\documentclass[fleqn,usenatbib]{mnras}

\usepackage{newtxtext,newtxmath}

\usepackage[T1]{fontenc}

\DeclareRobustCommand{\VAN}[3]{#2}
\let\VANthebibliography\thebibliography
\def\thebibliography{\DeclareRobustCommand{\VAN}[3]{##3}\VANthebibliography}

\usepackage{graphicx}	
\usepackage{amsmath}	

\title[Young stars in the relic NGC\,1277]{Lessons from the massive relic NGC\,1277: remaining in-situ star formation in the cores of massive galaxies}

\author[Salvador-Rusiñol et al.]{
N. Salvador-Rusiñol$^{1,2}$, \thanks{E-mail: nsalva@iac.es (NSR)}
A. Ferré-Mateu$^{1,2,3}$,
A. Vazdekis $^{1,2}$,
M. A. Beasley$^{1,2}$
\\
$^{1}$Instituto de Astrof\'{i}sica de Canarias, E-38200 La Laguna, Tenerife, Spain\\
$^{2}$Departmento de Astrof\'{i}sica, Universidad de La Laguna, E-38205 La Laguna, Tenerife, Spain\\ 
$^{3}$Institut de Ciencies del Cosmos (ICCUB), Universitat de Barcelona (IEEC-UB), E02028 Barcelona, Spain
}

\date{Accepted 2022 May 26. Received 2021 July 2}

\pubyear{2022}

\begin{document}
\label{firstpage}
\pagerange{\pageref{firstpage}--\pageref{lastpage}}
\maketitle

\defcitealias{SR19}{SR20}
\defcitealias{SR21}{SR21}

\begin{abstract}
Near-ultraviolet (NUV) spectroscopic studies have suggested that passively evolving massive, early-type galaxies host sub-one percent fractions of young stars in their innermost regions. We shed light on the origin of these stars by analysing NGC\,1277, a widely studied nearby prototypical massive compact relic galaxy. These are rare galaxies that have survived without experiencing significant size evolution via accretion and mergers since their formation at high redshift. We obtain a spectrum in the UV range within the central 1 kpc region of NGC\,1277. We compare a carefully selected set of optical and NUV line-strengths to model predictions with star formation histories characteristic of massive galaxies. We find a 0.8\% mass fraction of young stars in the centre of NGC\,1277, similar to that found in massive early-type galaxies. Given the limited accretion history of NGC\,1277, these results favour an intrinsic, in-situ, process triggering star formation at later epochs. Our results suggest a general constraint on the amount of young stars in the cores of massive early-type galaxies.  This amount should be assumed as an upper limit for the young stellar contribution in massive galaxies, as there might be present other contributions from evolved stars.
\end{abstract}




\section{Introduction}

There is growing evidence for a two-phase formation scenario for massive galaxies (e.g. \citealt{oser2010}). During the first phase, the cores of present-day massive early-type galaxies (ETGs) are formed dissipatively at z\,$>$\,2, while in the second phase the outer regions are the result of mergers and accretion. This is supported by a large number of studies that have shown that massive galaxies at z\,$>$\,2 are more compact than their local counterparts and have experienced significant growth both in size and, to a lesser degree, in mass (\citealt{trujillo2004, trujillo2006, vanderwel2008}). In this sense, massive elliptical galaxies increase on average by a factor of 4 in size and of 2 in stellar mass since z\,$\sim$\,2 to present (\citealt{trujillo2007}). Given the stochastic nature of galaxy evolution processes during the accretion phase, it is expected that a very tiny fraction of these massive compact galaxies have evolved until the present-day Universe preserving their structural properties and remaining untouched as compact and massive since their initial formation. These extremely rare and unique nearby objects are known as "relic" galaxies (\citealt{trujillo2014}). 

It has been recently shown with unprecedented precision that massive ETGs host sub-one percent fractions of young stellar populations (\citealt{SR19, SR21}, hereafter \citetalias{SR19} and \citetalias{SR21}, respectively), by fitting near-ultraviolet (NUV) and optical line-strength indices. Particularly, \citetalias{SR21} show that the young stellar populations in brightest cluster galaxies (BCGs), which are very massive galaxies located in the centres of galaxy clusters, are concentrated in their innermost regions ($<$\,2\,kpc). However, it becomes difficult to disentangle whether these young stars are formed due to intrinsic galaxy properties derived from the initial formation of the galaxy, or due to the experienced disruptions with surrounding galaxies and environment during the second phase. The exploration of the young stellar components in a massive relic galaxy would provide important clues of the possible origins for the formation of new stars in massive galaxies. If relic galaxies have recently formed stars, it would suggest that the main source of gas for the formation of these new stars is mainly associated with intrinsic galaxy processes, since these galaxies have not experienced important mergers with other galaxies since z\,$\sim$\,2. If, on the contrary, relics do not host any star formation activity, it would directly indicate that any finding of young stars in massive ETGs would be related to external processes.

To this end, in this work we investigate whether there are any young stars in the central region of NGC\,1277. NGC\,1277 is a low-redshift (z\,$\sim$\,0.0169) massive relic galaxy located in the Perseus cluster. It has been widely studied in terms of its morphology, stellar populations, kinematics, initial mass function (IMF) and super massive black hole (\citealt{vandenbosch2012, trujillo2014, martinnavarro2015b, ferremateu2017, yildirim2017}). It has a stellar mass of 1.2\,$\times\,10^{11}$\,M$_{\odot}$ and it is extremely compact, with an effective radius of only R$_{e}$\,=\,1.2\,kpc (\citealt{vandenbosch2012, trujillo2014}). Note that normal ETGs of the similar mass are on average four times larger. This indicates that NGC\,1277 has evolved until the present-day preserving its size without significant merging events since its initial epoch of formation \citep{beasley2018}. The bulk of its stellar populations is old and significantly enhanced in Mg (\citealt{trujillo2014, martinnavarro2015b, ferremateu2017}), indicating that most of its stars were formed in a extremely short and intense star formation burst at z\,$>$\,2, followed by passive evolution.  \citet{martinnavarro2015b} showed that this galaxy has a extremely bottom-heavy IMF at all radii, characteristic of the central regions of ETGs.

\section{Observations and data reduction} \label{sec:observations}

\subsection{NUV data}

We performed long-slit NUV spectroscopic observations of the relic galaxy NGC\,1277, carried out at the 2.5\,m Isaac Newton Telescope (INT), located at the Observatory del Roque de los Muchachos on La Palma (Spain). Observations were carried out in visitor mode on December 2018 under proposal program ID: 149-INT12/18B (PI: NSR). We employed the IDS spectograph using the R1200B grating centered at 3600\,\AA \ -- to acquire the bluest wavelength possible within the instrumental spectral range, 3100\,\AA \ -- with a x2 binning in both spatial and spectral direction. This corresponds to a spectral resolution of 3.5\,\AA. The INT slit was 3.3\,arcsec long and we opened it 2\,arcsec, positioning it along the major axis of NGC\,1277 (P.A: 95 deg). Total on-source time was 11\,h, with a varying seeing of 0.6\,--\,1.0\,arcsec.

\begin{figure}
\centering
	\includegraphics[width=1.\columnwidth]{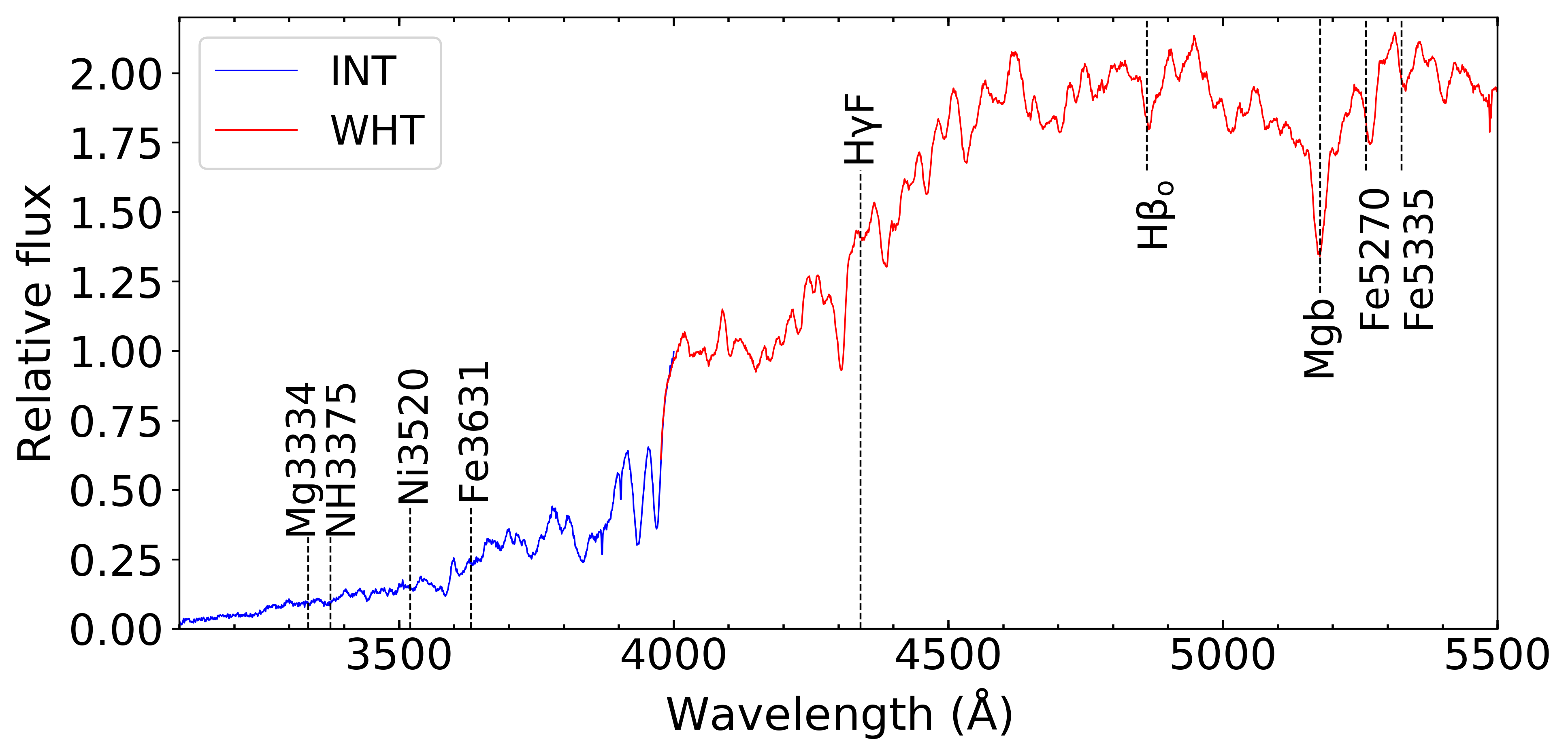}
    \caption{Innermost spectrum of NGC\,1277 covering the NUV and optical ranges, extracted within 1\,kpc from INT (blue) and WHT (red) spectra, joined after normalising at 4000\,\AA. Vertical lines indicate the central wavelengths of final set of NUV indices used for the fitting process in Section \ref{sec:fitting}. }
    \label{fig:relicsspec}
\end{figure}

We followed standard data reduction procedures using IRAF packages. These include bias subtraction, cosmic rays removal, flat-fielding, wavelength calibration and sky subtraction. Cosmic rays were removed with the python version of LA-cosmic package \citep{vandokkum2001}. A spectrum of NGC\,1277 was extracted within an aperture of 1\,kpc radius, to match the innermost aperture of the comparison work of \citetalias{SR21}. The extracted spectrum with useful information covers the spectral range 3100\,--\,4000\,\AA \ in the rest-frame and is shown in Figure \ref{fig:relicsspec}.

We could not flux calibrate the INT data but, fortunately, between 3100 and 4100\,\AA \ the instrumental sensitivity is quite flat. However, we investigated how this may affect the measured indices to find that there is a negligible difference ($\Delta$index\,$\sim$\,0.05 on average) in the NUV indices used when a given spectrum is not flux calibrated, since they cover a narrow spectral range that includes two pseudo-continua defined locally at both sides of each absorption feature, minimising the effect of a change in the continuum level and therefore, not affecting our results.

Before measuring NUV indices and comparing them with model predictions, we smoothed the observed spectrum and the SSP models to the same spectral resolution. We first obtained the kinematic broadening of the galaxy using \texttt{pPXF}. \citep{cappellari2004}. We use as templates the E-MILES SSP models \citep{vazdekis2016} and convolve the data to match the spectral resolution of the model template. The kinematics were derived using the wavelength range 3500\,--\,3950 \AA \ in rest-frame. We derive a stellar velocity dispersion of 350$\pm$8\,km\,s$^{-1}$ within 1\,kpc, which indicates that this is a very massive galaxy, and a redshift of z\,$\sim$\,0.0169, consistent values with \citet{trujillo2014} and \citet{yildirim2017}.

\subsection{Optical data}

The spectral range covered by the INT is missing information from the optical spectral range, which is needed to derive properties from the bulk of the stellar population. We thus use the published data obtained with the ISIS spectrograph in the 4.2\,m William Herschel Telescope (WHT). The reader is referred to \citet{trujillo2014} for a full description of the data. The optical data has a spectral resolution of 3.4\,\AA \ and reaches to 6200\,\AA \ with a very high quality. To match the NUV spectrum, we \ similarly extract the central 1\,kpc region and measure the stellar kinematics. The stellar velocity dispersion measured is 380$\pm$6\,km\,s$^{-1}$, slightly higher than the one we obtain from the INT data, mainly because of the different wavelength ranges used. However, the difference found in velocity dispersion between the INT and WHT spectra has a negligible effect in our measured indices. The optical spectrum is shown in Figure \ref{fig:relicsspec}.

\section{Methodology} \label{sec:methods}

In this section we describe the methodology employed to investigate the young stellar component in the centre of NGC\,1277. The method is based on that developed in \citetalias{SR19}. It consists of simultaneously fitting NUV and optical line-strengths indices with model predictions using different parametrisations of the star formation history (SFH) for massive galaxies.

\subsection{Stellar population models} \label{sec:sspmodels}

To analyse both the NUV and the optical indices we make use of the E-MILES single stellar population models described in \cite{vazdekis2016}. These models are based on empirical stellar libraries covering from the UV to the near-infrared wavelength ranges. The models are computed for different IMF shapes and slopes. In particular, we adopt a low-mass tapered "bimodal" IMF \citep{vazdekis1996}. The SSP models span an age grid from 63\,Myr to 14\,Gyr. For younger ages, we used the version of E-MILES models extended to 6.3\,Myr \citep{asad2017}, which are constructed with Padova isochrones \citep{girardi2000}. We select a metallicity grid that ranges from [M/H]\,=\,$-0.40$ to 0.22\,dex. Given the high metallicities obtained for NGC\,1277 in \citet{trujillo2014} and \citet{martinnavarro2015b}, we have performed a linear extrapolation with metallicity out to [M/H]\,=\,0.40.

\subsection{Analysis} \label{sec:fitting}

\begin{figure*}
\centering
	\includegraphics[width=1.\textwidth]{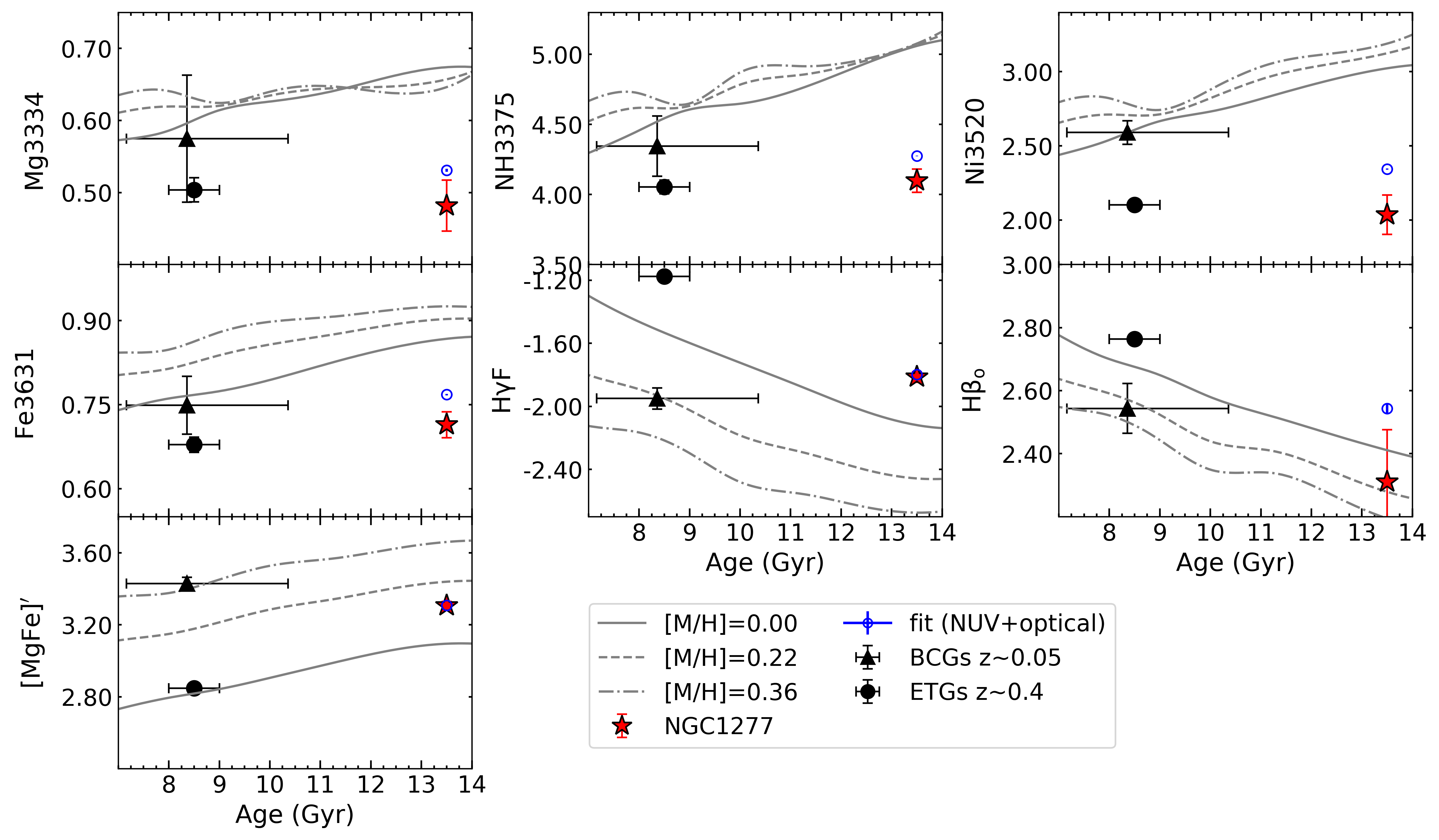}
    \caption{NUV and optical best-fitting line-strength indices. Each panel shows the selected NUV and optical SSP model indices as a function of age, for three different metallicities (solar, [M/H]\,=\,0.00 solid lines; [M/H]\,=\,0.22 dashed lines; [M/H]\,=\,0.36 dashed-dotted lines) for a bottom heavy bimodal IMF $\Gamma_{b}$\,=\,3.0. The observed measurements of NGC\,1277 are plotted as red stars as a function of the MLWA. The best-fitting model indices are plotted in blue. Measurements on the most massive ETG bin of \citetalias{SR19} are plotted as black circles. Values from the innermost bin from the BCGs studied in \citetalias{SR21} are represented as triangles. SSP models and data are at the same spectral resolution of 380\,kms$^{-1}$.}
    \label{fig:relicsindicesfits2}
\end{figure*}

Figure \ref{fig:relicsindicesfits2} shows the comparison between the observed and the model NUV indices for a single old SSP. The observations are plotted as a function of the mean luminosity-weighted age (MLWA), which is derived by fitting the observed optical indices H$\beta_{o}$ and [MgFe]$^\prime$ with SSP model predictions. Therefore, we assume an SSP to find the best-fitting solution of age and metallicity. These indices are senstive to the age and to the metallicity, respectively. The optical index [MgFe]$^\prime$ minimises the sensitivity to the [Mg/Fe] abundance ratio \citep{thomas2003}. The figure shows a discrepancy between the observed NUV indices of NGC\,1277 with model predictions. We will show that this mismatch is resolved by adding a small amount of recent star formation on top of an old stellar population. For comparison, the figure also shows the measurements of the most massive stacked spectrum ($\sigma$\,=\,300\,--\,340\,kms) of thousands of ETGs at z\,$\sim$\,0.4 studied by \citetalias{SR19}. We also show the index values of the innermost bin ($<$\,0.8\,kpc) of the stacked spectra of BCGs analysed in \citetalias{SR21}. Both data require a sub-one percent level of young stars on top of an old stellar population. Therefore, by comparing visually our measurements of NGC\,1277 to these data, we are able to predict that this massive compact relic galaxy will also require some small amount of young stars to match the observations.

We fit the NGC\,1277 line-strengths with model predictions derived from a parameterisation of the SFH typical for very massive galaxies, which considers a combination of an old and a young stellar population components. We refer the reader to \citetalias{SR19} for a full description of the modelling approach ({called \sl 1SSP+cSFR} model). For the old stellar component -- described by a SSP -- we assume prior distributions of ages from 1 to 14\,Gyr and metallicities ranging from --\,0.40 to 0.40\,dex. For this galaxy, we assume a bottom-heavy IMF of $\Gamma_{b}$\,=\,3.0, following \citet{martinnavarro2015b}. The young component is parametrised by assuming a constant star formation rate (SFR) in the last 1\,Gyr. The fitted stellar population parameters are the age and metallicity of the old stellar component, Age$\rm_{old}$ and [M/H]$\rm_{old}$, respectively, and the relative mass fraction of young stars formed during the most recent period (f$\rm_{1\,Gyr}$). 

The best-fit parameters are then estimated by comparing the observed line-strengths indices with those predicted by the model approach described above. Our index fitting method is based on a Markov Chain Monte Carlo (MCMC) algorithm, using the publicly available \textsf{emcee} routine \citep{foremanmackey2013} to fit simultaneously our set of indices. The best-fitting model is obtained by maximizing the log-likelihood function $\ln\mathcal{L}$. The best-fitting values are the median of each parameter distribution and the uncertainties are the 16-th and 84-th percentile levels, which are marginalized over the other parameters.

\subsubsection{Fits with NUV + optical indices}

We first investigate how each NUV index is affected by the young stellar component. For this purpose, we fit the optical indices, H$\gamma$F, H$\beta_{o}$ and [MgFe]$^\prime$ simultaneously with each NUV index, one by one. We find that almost all the indices provide similar best-fitting solutions, with a very old and metal-rich stellar component and a nearly similar contribution from young ($<$1\,Gyr) stars that ranges from 0.72\% to 0.88\%, in mass. Interestingly, indices with larger $\chi^{2}$ are those which were also discarded in \citetalias{SR21} due to their sensitivity not only to the young stellar component, but also to the alpha-element abundances to an even greater degree. Therefore, we base the following analysis on the NUV indices Mg3334, NH3375, Ni3520 and ﻿Fe3631, and optical indices simultaneously, which are highlighted in Figure \ref{fig:relicsspec}. We have tested the robustness of the present work by fitting this set of indices in the spectra of \citetalias{SR19}, where we used a larger set of indices. We find that we recover the same young mass fractions that were obtained with a larger, and bluer, set of indices than is used here.

We measure mass fractions for the young stellar component within the central 1\,kpc region of 0.8$\pm$0.06\%, on top of an old stellar component of an age Age$\rm_{old}$\,=\,12.6$\pm{0.7}$\,Gyr and metallicity [M/H]$\rm_{old}$\,=\,0.23$\pm{0.02}$\,dex. This indicates that the relic galaxy NGC\,1277 has experienced a low level of recent star formation in its central regions compatible with a specific SFR of $\rm \sim8\,x\,10^{-12}\,yr^{-1}$ within the last 1\,Gyr. The best-fitting indices are shown in Figure \ref{fig:relicsindicesfits2} and provide a better agreement with the observations with a combination of a small mass fraction in stars younger than 1\,Gyr, on top of an overwhelming dominating old component.

\subsubsection{Fits with NUV indices alone}

Because our INT data only covers the NUV range, missing the optical range, we also investigate how our results could be potentially affected if only fitting the NUV indices. We thus explore this effect by using the set of stacked spectra from BOSS, which are presented in \citetalias{SR19}. We repeat the fitting process in that data to see if we retrieve the amount of young stars obtained in \citetalias{SR19}, but, in this case only using indices defined within the spectral range of the INT data (i.e. Mg3334, NH3375, Ni3520, Fe3631). We find that relying on this set of indices alone we are not able to constrain small amounts of young stellar components. This indicates that it is required for optical indices to be also sensitive to the bulk of old stellar populations. We do not use other indices defined in the NUV spectral region since they are highly affected by the alpha-element abundance present in massive galaxies, as discussed in \citetalias{SR21}.

It is worth mentioning that without optical indices but using bluer indices down to Fe2609, \citetalias{SR19} were able to infer small amounts of young stellar populations on top of an old component. Note, however, that lower young mass fractions were obtained when optical indices are fitted simultaneously. Indices bluer than 3000\,\AA \ are much more sensitive to small fractions of young stars and can be used, with enough S/N, without the optical to study the recent star formation. On the contrary, NUV indices within the wavelength range 3100\,--\,3700 \AA \ not only are less sensitive to young components than bluer ones, but also less sensitive to the old stellar populations than the optical ones. This means that indices defined within this intermediate spectral range require optical indices to infer the young stellar components.

\subsubsection{Other possible contributions}

The presence of hot-evolved stars such as PAGBs stars might contribute also to decrease the NUV index strengths. However, as explored in \citetalias{SR19} and \citetalias{SR21} the best fits obtained with the PAGBs for the NUV indices translate to contributions in the optical range that are much higher than any expectation from the stellar evolution theory. Therefore, although PAGBs might be present, the major contribution in the NUV was attributed to the young stellar populations.

As for contributions of Horizontal-Branch (HB) stars from metal-poor stellar populations we have explored how the presence of these stars may impact on the NUV indices. For this purpose, we have combined two SSPs, one old and metal-rich, representative for the bulk of stars, and another old (13 Gyr) and metal-poor ([M/H] = -1.3), hosting a blue HB population. We find that the observed indices require a 20\% mass fraction of metal-poor stars to be approximately matched, and give a significantly worse (5 times larger) $\chi^2$. Such a mass fraction of metal-poor stars is much higher than what is expected for massive galaxies (e.g., \citealt{vazdekis1997, maraston2000}).

We also have performed additional modelling to explore the presence of contributions of extremely hot HB (EHB) from metal-rich stellar populations \citep{oconnell1999}. We use the evolutionary tracks for EHB stars (Cassisi priv. com.) and supersolar metallicity (as appropriate for NGC1277), which can be combined with the BaSTI models that are currently implemented in our E-MILES stellar population models. These EHB models have been computed by adopting the structural properties of a red giant branch (RGB) progenitor of 1 M$_\odot$ that reaches the RGB Tip at $\sim$13.5 Gyr. The following ZAHB (zero-age horizontal branch) and HB evolution has been computed by adopting the He core mass and surface He abundance at the RGB Tip and assuming an H-rich envelope mass of a few 10$^{-4}$ M$_\odot$. More details about these models can be found in \citet{cassisi2003} and \citet{pietrinferni2004}. We find that for a mass of 0.485M$_{\odot}$ the resulting EHB stars span from $\sim$7500 K to $\sim$20000 K with log(g) values ranging from $\sim$5 to $\sim$1.5. In order to implement these EHB models in E-MILES it was only necessary to change those stars that are more evolved than the RGB Tip in the models feeding our stellar population synthesis code. With this, we computed the corresponding supersolar SSP spectra with varying ages and IMFs to be confronted with our data.

We have fitted our observed indices with predictions from a combination of a standard old ($\sim$13 Gyr) and metal-rich ([M/H]=0.26) SSP with a EHB SSP model with similar age and metallicity. This has been made in the same way as we did with our standard models. The best fitting model requires a mass fraction of 42\% of the SSP with EHB stars. The observed indices are well fitted, and with very similar $\chi^{2}$ as achieved with the young contribution modelling, indicating that this kind of stars are able to match the NUV spectrum of this galaxy (as is the case of the young fraction). We also see that the SSP model with EHBs is much flatter than the
observed spectrum, as already shown in \citet{lecras2016}. Moreover, the EHB stars of the best fit model (with logarithmic (bimodal) IMF slope 3.0) contribute with $\sim$32\% to the total light in the U-band, i.e. roughly about 5 times the relative contribution of the standard SSP model (i.e. 6.5\%) for the same metallicity and age. The amount of these stars represent nearly the half of the HB stars in the combined populations. This fraction of stars is much higher than what is expected from multi-band HST NUV star counts studies in M31 and M32 (e.g. \citealt{brown1998}), and stellar clusters like wCen for which the EHB reaches 32\% \citep{dcruz2000}). These tests further confirm that the estimated fractions of young components must be taken as an upper limit, as other components such as old evolved stars might be also present and contributing in the NUV spectral range.

\section{Discussion and conclusions} \label{sec:discussion}

The old ages of nearby ETGs and the discovery of passively evolving massive galaxies at high redshift \citep{glazebrook2017} suggests that the star formation was efficiently ceased at earlier times. The majority of these high-redshift massive galaxies experienced a growth both in mass and size at later times leading to the massive and large ETGs we see in the present-day Universe.  Fortunately, without contamination from recent accretion \citep{beasley2018}, relic galaxies are the perfect targets to study the intrinsic regulation of star formation in massive galaxies. We have performed the first study concerning the recent star formation in the prototype relic galaxy, NGC\,1277. For this purpose, we measured the NUV indices and found that their line-strengths deviate significantly with respect to the model predictions for an old stellar population. This mismatch is resolved by adding a small fraction of young stellar populations. The best-fitting solutions estimate that a 0.8\% in mass of stars within 1\,kpc is formed in the most recent 1\,Gyr of its SFH. The SFHs inferred by previous authors for this galaxy, however, did not evidence recent star formation, possibly due to the relative insensitivity of the full-spectrum fitting in the optical range to such small mass fractions of young stars. We will address in a forthcoming paper (Ferré-Mateu et al. in prep) the capability to infer small amounts of young stellar components in massive galaxies from different spectral regimes by using the full-spectrum fitting. 

In addition, we have also fitted the observations with employing SSPs with hosting EHBs. We find that the indices of NGC\,1277 in the U-band spectral range can also be reproduced by a combination of old metal-rich SSPs where a mass fraction of 42\% corresponds to a component with EHB stars with temperatures in the range 7000 -- 20000 K. However, with the UV spectral range covered here (U band) and the methodology employed in this study we are not able to distinguish between the EHB and young mass-fractions scenarios. It is worth to investigate additional constraints that allow to disentangle the contrubution in the NUV of massive galaxies from young stellar populations and from hot evolved stars.  We aim at addressing this issue with the appropriate data and modelling in forthcoming works. Therefore, we further remark that the young mass-fractions provided here must be interpreted as upper limits, as other stellar components such as HB branch stars or PAGB stars may be also contributing to the NUV indices.

\begin{figure}
\centering
	\includegraphics[width=1.\columnwidth]{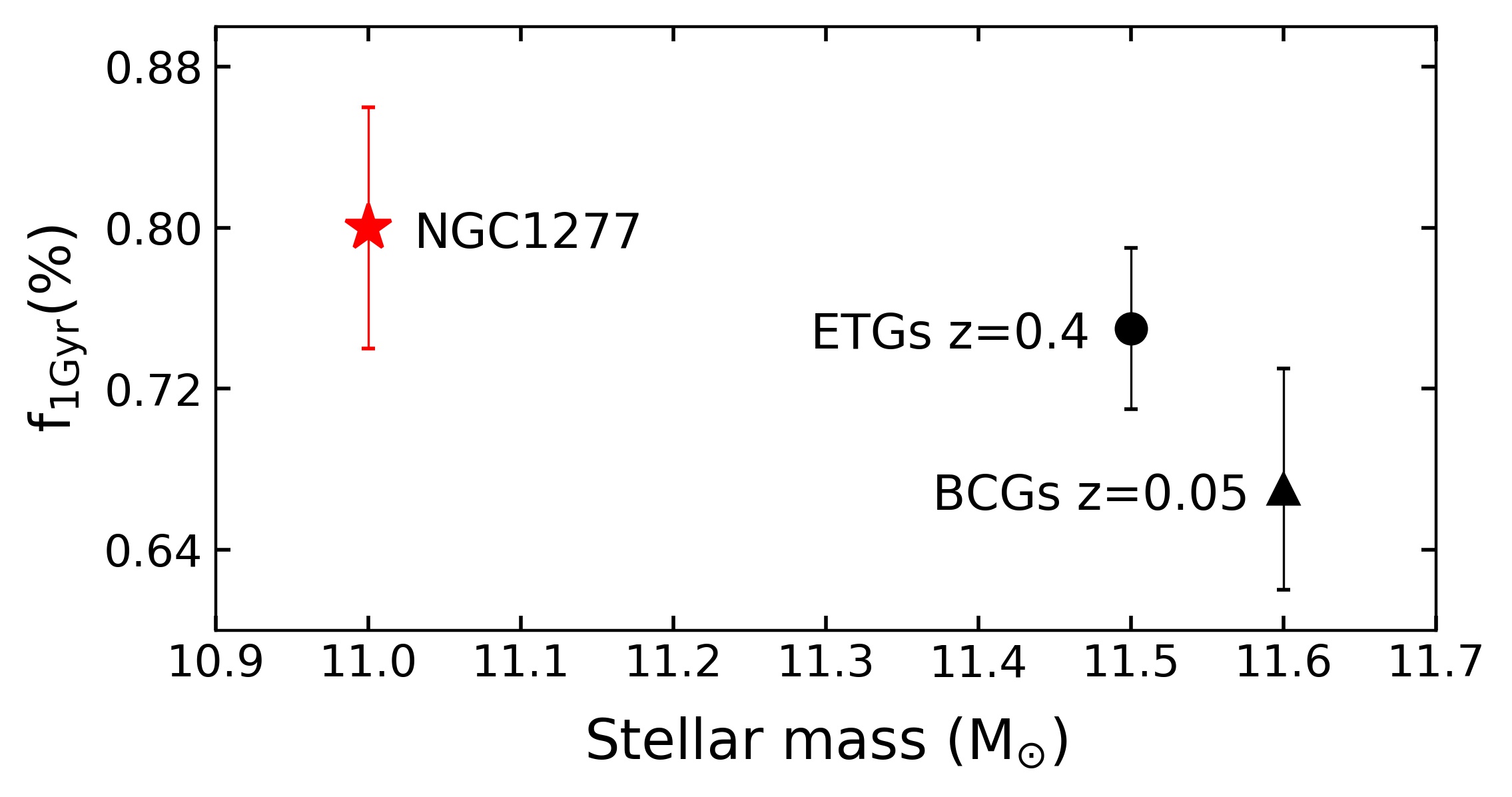}
    \caption{Amount of young stars ($<$\,1\,Gyr) within 1\,kpc central region as a function of the galaxy stellar mass. NGC\,1277 is shown as a red star, the massive ETG population at z$\sim$0.4 studied in \citetalias{SR19} is indicated as a black circle and the average from 6 nearby BCGs analysed in \citetalias{SR21} is shown as a black triangle. }
    \label{fig:comparison}
\end{figure}

To put our results for NGC\,1277 into a broader context, we show in Figure \ref{fig:comparison} the amount of young ($<$\,1\,Gyr) stars formed within the innermost 1\,kpc region as a function of the galaxy stellar mass for the relic NGC\,1277 and for the massive ETG population at z\,$\sim$\,0.4 and the BCGs at z$\sim$0.05 according to \citetalias{SR19} and \citetalias{SR21}, respectively. Note that we show the estimates for a similar aperture of 1\,kpc, to make a fair comparison between the three samples. NGC\,1277 is the only individual galaxy shown in the figure, while the other results represent an average young mass fractions from thousands of massive ($\sigma_0\,>$\,300\,kms$^{-1}$) ETGs and six nearby BCGs. Due to the high stellar velocity dispersion, the massive ETGs population is expected to be located in the centres of galaxy clusters. Massive ETGs and BCGs, which have evolved by accreting smaller satellite galaxies during the accretion phase, also have a sub-one percent level of young stars ($\sim$\,0.7\% in mass) in their central regions. These amounts are very similar to the value found in the relic galaxy. This suggests that \textit{these low levels of young stellar populations in the cores of massive ETGs come from self-regulated gas returned to the interstellar medium from stellar evolution}. Such a scenario would imply that the cores of massive galaxies evolve in a very similar way as the relic galaxies do as derived from NGC\,1277.

NGC\,1277 is also exceptionally interesting for hosting one of the most-massive black holes known (\citealt{vandenbosch2012, ferremateu2015}). AGN suppression of star formation is invoked in almost every state-of-the-art numerical simulation. Feedback from energetic jets of AGN in massive galaxies is thought to heat and expell the available cold gas, preventing the star formation at high redshift. The AGN feedback took place mainly at high-redshift (z\,$>$\,2), but the mechanisms that keep massive galaxies quenched from earlier times to the local Universe remain unknown. Observationally, \citet{martinnavarro2018a} suggested that this quenching is strongly linked to the mass of the black hole. The overly massive black hole of NGC\,1277 makes it one of the most AGN-dominated galaxies in the nearby Universe. Thus, it would be expected that NGC\,1277, with its supermassive black hole, has lower levels of star formation than ETGs for the same galaxy mass. However, our results show that NGC\,1277 has similar fractions of young stellar components than the massive ETG population. This indicates that at the present epoch the central supermassive black hole is playing no significant role in regulating the star formation in NGC\,1277. By extension, our results suggest that AGN is not directly regulating the star formation in massive, passively evolving ETGs at z\,=\,0.

Because NGC1277 is located in the Perseus cluster, a very dense gallaxy cluster, one possibility would be that the hot halo structures surrounding the galaxy prevented gas to get sufficiently cold, triggering the star formation. In addition, NGC\,1277 is likely a satellite of the large and massive BCG NGC\,1275, which is presently accreting large amounts of gas (\citealt{conselice2001, lim2008}) and possibly acts as an attractor, preventing further accretion onto NGC\,1277. Therefore, studying the young stellar populations in relic galaxies located in lower-density environments, i.e. more isolated, like the massive compact galaxies MRK\,1216 and PGC\,32872 \citep{ferremateu2017}, will be relevant for disentangling quenching mechanisms at low-redshifts.

Interestingly, the amount of gas returned into the interstellar medium from dying stars is mostly regulated by the IMF. NGC\,1277 seems to have a nearly constant IMF with rather bottom-heavy shape at all radii. A bottom heavy IMF would produce an amount of gas from stellar evolution that would be sufficient to form the estimated mass fraction of young stars in NGC\,1277. Therefore, it would be highly interesting to explore how these young stars distribute within relic galaxies, and see whether they are also concentrated in the innermost regions as it has been recently shown for the BCG sample studied in \citetalias{SR21}. The study of young stellar populations in relic galaxies therefore gives unique insights into the formation and evolutionary processes of the most massive galaxies.

\section*{Acknowledgements}

We thank Santi Cassisi for providing us their newly developed evolutionary tracks. We are grateful to John Beckman for our interesting discussions. NSR acknowledges funding from Spanish Ministry of Science, Innovation and Universities (MCIU), through research project SEV-2015-0548-16-4 and predoctoral contract BES-2016-078409. AFM has received financial support through the Postdoctoral Junior Leader Fellowship Programme from 'La Caixa' Banking Foundation (LCF/BQ/LI18/11630007). MAB and AFM acknowledge support from the Severo Ochoa Excellence scheme (SEV-2015-0548 and CEX2019-000920-S). All authors acknowledge support from grant PID2019-107427GB-C32 from the MCIU.

\section*{Data availability} 

This work is based on observations made with the Isaac Newton Telescope at the Observatorio del Roque de los Muchachos (La Palma, Spain) under programme ID 149-INT12/18B (PI: NSR). The spectra analysed that supports the plots within this paper are available from the corresponding author upon request. The E-MILES SSP models are publicly available at the MILES website (http://miles.iac.es).



\bibliographystyle{mnras}
\bibliography{references} 




\appendix

\section{Some extra material}

If you want to present additional material which would interrupt the flow of the main paper,
it can be placed in an Appendix which appears after the list of references.


\bsp	
\label{lastpage}
\end{document}